# SPECTRAL THEORY FOR DISSIPATION MECHANISM

# OF WIND WAVES

Vladislav G. Polnikov[1]

[1]A.M. Obukhov Institute for Physics of Atmosphere, Russian Academy of Sciences

Pyzhevskii lane 3, 119017, Moscow, Russia,

E-mail: polnikov@mail.ru

Tel: +7-916-3376728

Fax: +7-495-9531652

April 2010

ABSTRACT

A systematic and full description of the theory for a dissipation mechanism of wind wave energy in a spectral representation is given. As a basis of the theory, the fundamental is stated that the most general dissipation mechanism for wind waves is provided by the viscosity due to interaction between wave motions and turbulence of the water upper layer. The latter, in turn, is supposed to be induced by the whole aggregate of dissipation processes taking place at the air-sea interface. In the frame of phenomenological constructions of nonlinear closure for Reynolds stresses, it is shown that the dissipation function is generally a power series with respect to wave spectrum, starting from a quadratic term. Attracting previous results of the author, a simplified parameterization of the general theoretical result is done. Physical meaning for parameters of the dissipation function and its compliance with the new experimental facts established in this field for the last 5-10 years is discussed. Summarized theoretical results were verified in the mathematical shells of the well known numerical models for wind waves, WAM and WAEWATCH. Evidence is given, illustrating a superiority of the proposed model modifications. Prospects for elaboration of the theory are discussed.

1.  **Introductory analysis and pose of the problem**

As well known, wind-sea is a picturesque but rather complicated and dangerous phenomenon taking place at the air-sea interface each time a remarkable wind is present. For the reasons of easy visibility, complexity, and danger feature of the phenomenon, its study has more then a century-long history. Herewith, a lot of problems have not been solved in this field till now. One of them is construction of physical model and description of mechanism for wind wave energy dissipation. This paper is devoted to precipitate solution of this problem.

From mathematical point of view, a wind wave field has a stochastic feature, and its properties should be governed by a proper statistical ensemble. Therefore, the best way of the phenomenon description lies in the domain of statistical characteristics. The main of them is the two-dimensional spatial wave energy spectrum, $S(\mathbf{k},\mathbf{x},t) \equiv S$, spread in the space, $\mathbf{x}$, and time, $t$. The space-time evolution of this characteristic is described by the so called transport equation written in a spectral representation (Komen et al. 1994) of the kind

$$\frac{\partial S}{\partial t} + C_{gx} \frac{\partial S}{\partial x} + C_{gy} \frac{\partial S}{\partial y} = F \equiv NL + IN - DIS \ . \tag{1}$$

Here, the left-hand side is the full time-derivative of the spectrum, and the right-hand side is the so called source function ("forcing"), $F$. Vector ($C_{gx}$, $C_{gy}$) is the group velocity one corresponding to a wave component with wave vector $\mathbf{k}$, which is defined by the ratio

$$\mathbf{C}_g = \frac{\partial \sigma(k)}{\partial k} \frac{\mathbf{k}}{k} = ( C_{gx}, \ C_{gy} ) \ . \tag{2}$$

Dependence of frequency on the wave vector $\sigma(\mathbf{k})$ is given by the expression

$$\sigma = \sqrt{gk} \tag{3}$$

known as the dispersion relation for the case of deep water, considered below.

The left-hand side of equation (1) is responsible for "mathematical" part of a model, which is not discussed here, whilst the physical essence of a model is held by the source function, $F$. At present, it is wide recognized that $F$ can be written as a sum of three terms – three parts of the united evolution mechanism for wind waves:

- The rate of conservative nonlinear energy transfer through a wave spectrum, $NL$, ("nonlinear-term");

- The rate of energy transfer from wind to waves, $IN$, ("input-term");

- The rate of wave energy loss due to numerous dissipative processes, $DIS$, ("dissipation-term").

A difference of mathematical expressions (parameterizations) for the terms of source function defines a physical specification of each certain model. The worldwide spread models, WAM (The WAMDI group 1988) and WAWEWATCH (WW) (Tolman and Chalikov 1996), are the representatives of such a kind models.

Note that equation (1) has a meaning of the energy conservation law applied to each spectral component of wave spectrum, and for this reason Eq. (1) could be postulated phenomenologically. Nevertheless, under some simplifications, the general kind of Eq. (1) and each separate term of source function $F$ can be derived from the basic hydrodynamic equations of the kind

$$\rho \frac{d\mathbf{u}}{dt} = -\vec{\nabla}_3 P - \rho \mathbf{g} + \mathbf{f}(\mathbf{x}, t); \Big|_{z = \eta(\mathbf{x}, t)} \tag{4}$$

$$\frac{\partial \rho}{\partial t} + \vec{\nabla}_3 (\rho \mathbf{u}) = 0 \tag{5}$$

$$u_z \Big|_{z = \eta(\mathbf{x}, t)} = \frac{\partial \eta}{\partial t} + (\mathbf{u} \vec{\nabla}_2 \eta) \tag{6}$$

$$u_z \Big|_{z = -\infty} = 0 \tag{7}$$

where the following designations are used: $\rho(z,t)$ is the fluid density; $\mathbf{u}(\mathbf{x},z,t) = (u_x, u_y, u_z)$ is the velocity field; $P(\mathbf{x},z,t)$ is the atmospheric pressure; $g$ is the acceleration due to gravity; $f(\mathbf{x},z,t)$ is the external forcing (surface tension, wind stress and so on); $\eta(\mathbf{x},t)$ is the surface elevation field; $\mathbf{x} = (x,y)$ is the horizontal coordinates vector; $z$ is the vertical coordinate up-directed; $\vec{\nabla}_2 = (\frac{\partial}{\partial x}, \frac{\partial}{\partial y})$ is the horizontal gradient vector; $\vec{\nabla}_3 = (\vec{\nabla}_2, \frac{\partial}{\partial z})$ is the full gradient, and the full time-derivative operator is defined as $\frac{d}{dt}(...) = \left( \frac{\partial}{\partial t} + \mathbf{u}\vec{\nabla}_3 \right)(...)$.

Going to the problem pose, let us remind shortly the most important results in this field.

All the summands of source function $F$ have been derived already to some extent from equations (4)-(7) (Cavaleri et al. 2007, Komen et al. 1994). Among them, the nonlinear term $NL$ is the best theoretically justified. As was shown for the first time in (Hasselmann 1962) and later confirmed in (Zakharov 1968, 1974), this term (under some simplifications) can be expressed in the form

$$NL(S) \equiv F3(S) \qquad (8)$$

where $F3(S)$ is a very complicated integral–kind functional of the third power in $S$, exact expression of which is not needed here. Properties of this term are studied rather well, and the most effective parameterization of $NL$ has been constructed and verified long ego (see details and references in Polnikov and Farina 2002). This point is not the object of the paper, and we will not dwell more on this term.

Input term $IN$ is not so well studied, though all principal features of it are more or less known. A worldwide recognized form of $IN$, corresponding to the pioneer model of Miles (1957), has the following general kind

$$In = \beta(\sigma,\ \theta,\ \mathrm{u*}, \theta_\mathrm{w})\sigma S(\sigma,\ \theta) \qquad (9)$$

where $\beta(\sigma, \theta, u_*, \theta_w)$ is the wave growing increment representation of which is the most problematic. Arguments of this function are as follows: $\sigma$ and $\theta$ are the frequency and direction of propagation of a proper spectral wave component; $u_* = \left(C_d(z)W^2(z)\right)^{1/2}$ is the friction velocity defined from the wind speed at a standard horizon, $W(z)$, via especially calculated drag coefficient, $C_d(z)$; and $\theta_w$ is the mean wind direction. Due to complexity of the system under consideration, detailed presentation of $\beta(...)$ can not be derived theoretically form the system of Eqs. (4)-(7). Therefore, in numerical models they use different kinds of phenomenological approximations for it, having features corresponding to ones found in empirical measurements and numerical simulations (Cavaleri et al. 2007, Chalikov and Sheinin 1998, Komen et al. 1994). Representation of *IN*(*S*) in the form (9) is sufficient for us, to get main aim of the paper, and therefore we will not give here more details.

Finally, consider very shortly the present state of our knowledge about dissipation term *DIS*.

First of all, we should state that there is not any widely recognized spectral representation for function *DIS(S)* (see the most recent review for a state-of-the-art in Cavaleri et al. 2007). In particular, there is not understanding what is the power of spectrum in function *DIS(S)*: the first power (as used in WAM and WW according to Hasselmann's theory 1974), the second power (as proposed in Polnikov 1993, 1995, 2005), or any higher powers (according to Phillips 1985, Donelan 2001, Hwang and Wang 2004).

Second, there is not any theory where the issue of wind wave dissipation is considered from the most general point view. All of innumerous theoretical papers, including direct derivations (Hasselmann 1974), dimensional considerations (Phillips 1985), numerical simulations ( Zakharov et al. 2007), all of them do not take into account a presence of small-scale turbulence in the water upper layer, which interacts strongly with waves in reality. All present theoretical considerations are mainly restricted by different aspects of wave breaking processes or accompanying effects, like a white-capping in Hasselmann (1974).

Besides earlier papers by the author (Polnikov 1993, 1995, 2005), the only exclusion is the paper by Tolman and Chalikov (1996) where these authors tried to estimate the role of turbulent viscosity $\nu_T$ in the wave energy dissipation. But they have restricted themselves with a very certain kind of parameterization for the viscosity coefficient what have led them to a very particular representation of $\nu_T(S)$, vulnerable from a theoretical point of view.

The author's earlier papers mentioned above, dealing with the dissipation of wind wave due to turbulence, are also rather particular and unsystematic. Thus, the problem of construction of a more general and logically selfconsistent theory is strongly in demand.

Third, some words about experimental results. Here we will not dwell on review of experimental research in this field, as far as it is well done in the recent paper (Babanin 2007). It needs only to mention that, in our mind, any more or less realistic experimental study of the dissipation mechanism for wind waves are hardly possible, especially in a spectral presentation. That is due to presence a lot of invisible and immeasurable processes in the water upper layer, governing the dissipation of surface waves. For this reason the most of empirical measurements are restricted by study of different aspects of wave breaking process, only.

We would say more. In our mind, the breaking process is not directly a dissipation of waves. It is rather a fast spreading of energy among sideband wave scales. And the experimental measurement of Young and Babanin (2006) directly demonstrates it (see Fig. 1 from the reference). Thus, we state that measurement of dissipation rate due to the breaking dose not correspond totally to a real dissipation rate in a spectral representation.

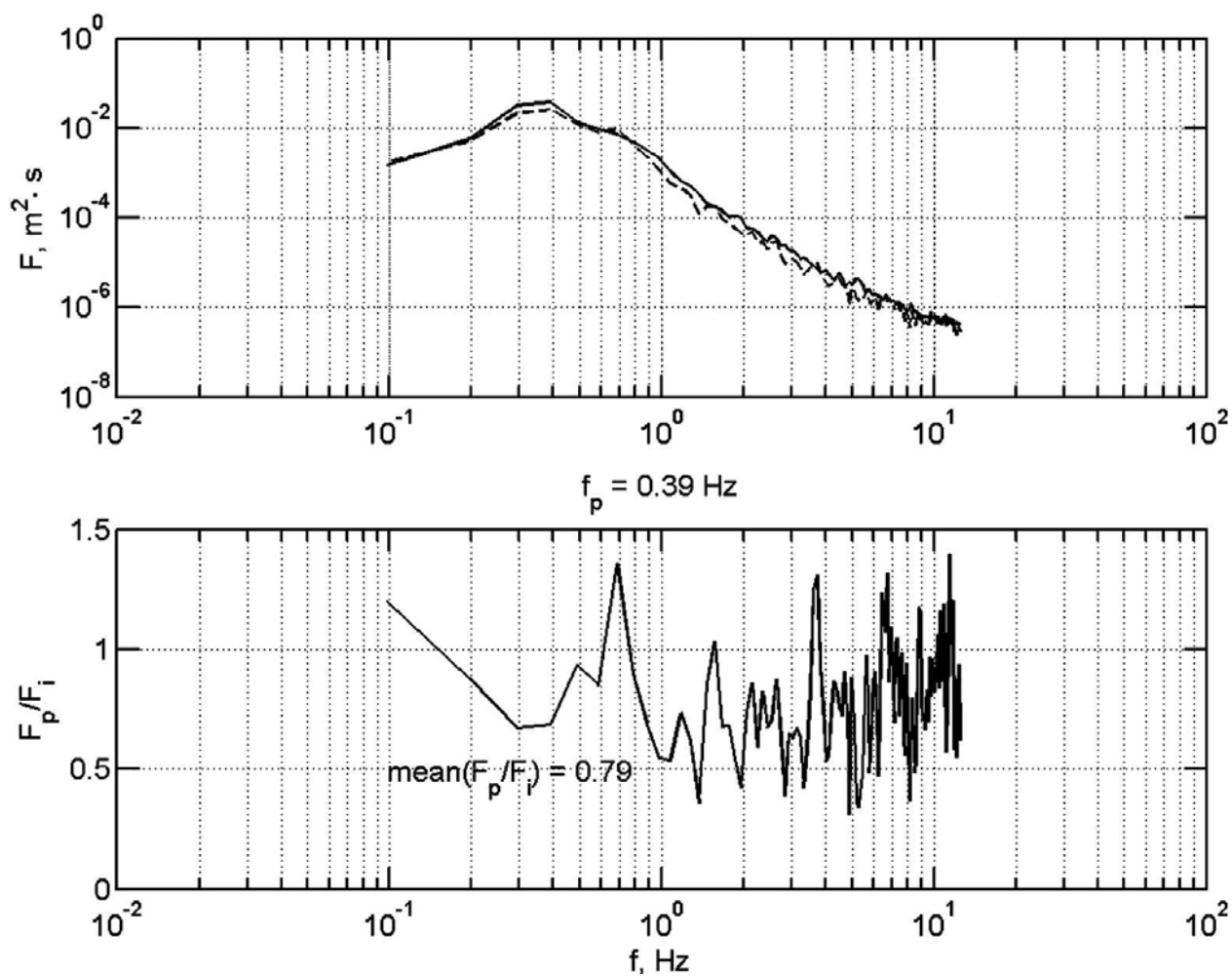

Fig. 1. Top panel: Mean power spectrum of incipient-breaking (solid line) and postbreaking (dashed line) waves. Bottom panel: Ratio of the spectra shown in top panel (following to Young and Babanin 2006) .

Nevertheless, for the last 5-10 year, the efforts of experimenters are rather fruitful, and some established empirical effects related to the wind-wave breaking process are quite interesting. In particular, following the papers (Banner and Tiang 1998, Babanin et al. 2001) and some others (see references in Young and Babanin 2006), we have to mention the following empirical effects related to the wave breaking process:

E1) Threshold feature of the wave breaking;

E2) Influence of long wave breaking on intensity of breaking for shorter waves;

E3) Different features of the breaking rate for dominant waves and for waves in the tail part of wave spectrum;

E4) More intensive breaking for waves running at some angle to the mean wind direction (two-lobe feature of the angular function for breaking intensity).

All these effects can serve as guidelines, to get a true theoretical representation of *DIS(S)*.

Present paper is directly devoted to construction the most general and logically self consistent theory for dissipation mechanism of wind wave. The main fundamentals of general theory, main equations, derivations, and approximations are given in Sec. 2. Specification for the main result of general theory, description of physical meaning for parameters of the derived dissipation function, and comparison the features of the latter with the main experimental and numerical results in this field (effects E1-E4 mentioned above) are presented in Sec. 3. In the same section, in addition, it is shown the evidence of successful implementation of the theoretical result into numerical simulations of wind waves. In the final Sec. 4, the domain of validity of the proposed dissipation mechanism for description of wind wave spectrum evolution is discussed, and the prospects for further development of the theory are pointed out.

## 2. The main theory for wind wave dissipation

### 2.1. Basic statements.

The main fundamental of the proposed theory states that on the scales of Eq. (1) validity, the most general reason for wind wave dissipation is a turbulence of the water upper layer. Herewith, specification of processes producing the turbulence is quite unprincipled.

Really, it is evident that a reasonable part of the turbulence intensity is provided by the wave breaking processes. Thus, the breaking processes are taken into account in our statement, though by an implicit way. Herewith, it is equally clear that accompanying processes, as like as sprinkling, white capping, shear currents and wave orbital motions instabilities (taking place both in an atmosphere and in a water boundary layer), air-bubble clouds, and so on, all of them, in terms of hydrodynamics, are chaotic motions without any determined scale, i.e. the turbulence motions. Contribution of these motions to the wind-wave dissipation is to be taken into account. In other words, the proposed theoretical approach is based on including into considerations all dissipative processes leading to production of turbulence in the water upper layer.

According to the said, without any restriction of the consideration generality, the current field in a waving water layer can be written in the form of two constituents

$$\mathbf{u}(\mathbf{x}, z, t) = \mathbf{u}_w(\mathbf{x}, z, t) + \mathbf{u}'(\mathbf{x}, z, t) \ . \tag{10}$$

The first summand, $\mathbf{u}_w$, in the right-hand side (further, the r.h.s.) of (10), we treat as the potential motion attributed to wind waves. This motion is governed by the system of equations corresponding to one given by Eqs. (4)-(7), but written in the potential approximation. The second summand, $\mathbf{u}'$, is treated as the turbulent constituent of full velocity, totally uncorrelated with $\mathbf{u}_w$ in statistical sense.

As regards to surface elevation $\eta(\mathbf{x}, t)$, we do not introduce analogous decomposition, attracting the well known Hasselmann's hypothesis of "a small distortion in mean" for the surface profile (Hasselmann 1974). This allows us to use conception of surface elevation field $\eta(\mathbf{x}, t)$ in a traditional, commonly used sense.

### 2.2. *Reynolds stresses.*

Following to the said, let us rewrite basic equations (4) and (6) without any external force in the standard tensor kind

$$\frac{\partial u_i}{\partial t} + \sum_j u_j \frac{\partial u_i}{\partial x_j} = -g\,\delta_{i,3}\Big|_{z=\eta(\xi, t)} \qquad , \tag{11}$$

$$\frac{\partial \eta}{\partial t} = u_3 - \sum_{i=1,2} u_i \frac{\partial \eta}{\partial x_i}\Big|_{z=\eta(\xi,t)} \qquad . \qquad (12)$$

Here, indexes $i, j$ take values 1, 2, 3, corresponding to two horizontal and one vertical coordinates, respectively, and the sub-index, $\Big|_{z=\eta(\xi,t)}$, means that the both equations are written at the interface. Further the latter sub-index will be omitted for simplicity.

Substitution of ratio (10) into Eqs. (11) and (12), with a successive averaging of them over the turbulent scales and accounting the condition of incompressibility, $\sum_j \partial u_j / \partial x_j = 0$, and absence of a fluctuation part for $\eta(\mathbf{x}, t)$, permits us to write the following system of equations for the wave motions

$$\frac{\partial \overline{u_i}}{\partial t} + \sum_j \overline{u}_j \frac{\partial \overline{u_i}}{\partial x_j} = -g\delta_{i,3} - \sum_j \frac{\partial < u_i' u_j' >}{\partial x_j}, \qquad (13)$$

$$\frac{\partial \overline{\eta}}{\partial t} = \overline{u}_3 - \sum_{i=1,2} \overline{u}_i \frac{\partial \overline{\eta}}{\partial x_i} \qquad . \qquad (14)$$

Here, the mean wave variables, $\overline{\eta}$ and $\overline{u}_i$, are denoted with the bar which will be omitted later with the aim of simplicity, and the additional term in the r. h. s. of Eq. (13), appearing due to nonlinearity of the system, is

$$\sum_j \frac{\partial < u_i' u_j' >}{\partial x_j} \equiv P_i \qquad . \qquad (15)$$

Physically, this term represents disturbing force $\mathbf{P}$ providing for the wave motion dissipation. To convince in this, it is enough to accept a simplest parameterization of the force in terms of wave velocity, $\mathbf{u}$, of the kind

$$P_i \equiv \sum_j \frac{\partial < u_i' u_j' >}{\partial x_j} = -\nu_T \sum_j \frac{\partial^2 u_i}{\partial x_j^2} \qquad (16)$$

with a constant value of factor $\nu_T$. In such a case, term (16) is absolutely equal to expression for the typical molecular viscosity force, what, by solution of the system (13)-(14) in a linear approximation and in a spectral representation for variables $\eta$ and $\mathbf{u}$, leads to the well known equation for the temporal evolution of wave spectrum of the kind (Hasselmann 1960)

$$\partial S(\mathbf{k},t) / \partial t = -2\bar{\nu}\sigma_k S(\mathbf{k},t), \qquad\qquad (\bar{\nu} = \nu_T k^2 / \sigma_k) \qquad\qquad (17)$$

having an exponentially decay solution.

Taking into account that the introduced constant $\nu_T$ has dimension of viscosity, and the solution of (17) has a decay feature for a wave spectrum, with no doubt one can state that the additional term (15) in Eq. (13) has meaning of the dissipative term provided by turbulence in the water upper layer. This simple consideration helps us to draw very important consequences.

Really, the numerator under the sum in expression (15) is a very well known magnitude in the turbulence theory, which is called the Reynolds stress (Monin and Yaglom 1971)

$$< u_i^{'} u_j^{'} > \equiv \tau_{ij} \qquad\qquad (18)$$

(here it is normalized by the water density). Methods of parameterization for $\tau_{ij}$ are well developed. Therefore, to complete the theory, it needs to specify a representation for $\tau_{ij}$ in terms of wave variables, $\eta$ and $\mathbf{u}$, to find evolution equation of the form (17), and to ascribe to the r.h.s. of final equation the physical meaning of dissipation term, *DIS*. To get final result, it is reasonable to use some principles of the Hasselmann's approach to this problem (Hasselmann 1974). Just this program will be realized below.

### 2.3. Phenomenological closure of Reynolds stress

First of all, let us formulate the main grounds of our concept for a procedure of Reynolds stress closure, the aim of which is to express the turbulent characteristic $\tau_{ij}$ via the wave variables $\eta$ and $\mathbf{u}$. One of them is the hypotheses of "a small distortion in mean", mentioned above. This hypotheses permits to safe commonly used conception of wave profile $\eta(\mathbf{x},t)$ and introduce all derivatives of $\eta(\mathbf{x},t)$, if needed.

The second ground is an assumption that a relative value of Reynolds stress term $P_i$ is much greater of the dynamical nonlinear term described by the second summand, $\sum_j u_j \dfrac{\partial u_i}{\partial x_j}$, in the l. h. s. of Eq. (13). Thereby we postulate the statement of "strong" nonlinearity of the turbulent type, permitting to neglect the dynamical nonlinearity in this problem. Thus, we should solve, in fact, the following system of equations

$$\frac{\partial u_i}{\partial t} + g\delta_{i,3} = -P_i(\mathbf{u},\eta) \qquad , \qquad (19)$$

$$\frac{\partial \eta}{\partial t} = u_3 \qquad . \qquad (20)$$

Third. The kind of closure (16) is too much simple, reflecting the meaning of "forcing" term (15) qualitatively, only. The principle shortage of this closure is reducing the nonlinear dynamics to the linear one. It is evident that more complex, nonlinear closure of the forcing term is more adequate to physics of the processes considered. The problem is to find such a closure which would be more general and have a reasonable physical treatment. Below we try to do it.

One of such a kind version of the stress closure is related to using the concept of Prandtl mixing length, $L$, allowing to express turbulent fluctuation velocity $\mathbf{u}'$ via derivative of wave velocity field $\mathbf{u}$ in the form (Monin and Yaglom 1971)

$$u_i' = \lambda_i \partial u_i / \partial x_i \qquad (21)$$

where $\lambda_i$ is the so called "length of turbulent mixing" having a stochastic feature. In such a case, Reynolds stress becomes a nonlinear function of wave variables

$$\tau_{ij} = <\lambda_i \lambda_j (\partial u_i / \partial x_i)(\partial u_j / \partial x_j)> = L_{ij}(\partial u_i / \partial x_i)(\partial u_j / \partial x_j), \qquad (22)$$

what changes radically both the structure of final dynamic equations and solution of them.

Note that closure (21), related to spatial derivatives of wave velocity $\mathbf{u}$, is quite adequate for a horizontally homogeneous, near-wall turbulence. In such a case, values $L_{ij}$ are ascribed to spatial scales of turbulent eddies, magnitudes of which are to be postulated. In our case, the wave induced turbulence is realized under conditions of waving interface, therefore, some modifications of the approach are needed. In particular, in the case of waving interface, it is very reasonable to use closure versions related to derivatives of wave profile $\eta(\mathbf{x},t)$. For example, we might use expressions of the kind

$$u_i^{'} = \vartheta_i \partial \eta / \partial x_i \qquad (23)$$

where values $\vartheta_i$ (by analogy to $\lambda_i$) have a meaning of a random "mixing velocity".

Summarizing all proposals postulated above, it is quite justified to represent the forcing term in the following rather general form

$$-P_i(\mathbf{u},\eta) = < \sum_j \frac{\partial}{\partial x_j} \{ \left[ \lambda_i(\partial u_i / \partial x_i) + \vartheta_i(\partial \eta / \partial x_i) \right] \left[ \lambda_j(\partial u_j / \partial x_j) + \vartheta_j(\partial \eta / \partial x_j) \right] \} > . \qquad (24)$$

Closure (24) maintains the following principal features of the problem:

(a) Nonlinear nature of the dissipation process;

(b) Dependence of the turbulent forcing on gradients of both surface elevation field $\eta(\mathbf{x},t)$ and velocity field $\mathbf{u}(\mathbf{x},z,t)$.

Here we should state that further specification of coefficients $\lambda_{i,j}$ and $\vartheta_{i,j}$ in form (24) is not principal at the moment. Moreover, as far as we do not know real processes generating turbulence of the water upper layer, there is no sense to construct any more complicated and detailed approximation for forcing term in the physical space, i.e. $P_i(\mathbf{u},\eta)$ (as they have been done in earlier papers by the author, Polnikov 1993, 1995). At present stage of the theory derivation, it is the most important to take account the nonlinear feature of forcing term, only. As it will be shown below, this fact itself gives sufficient grounds for a further finding a general kind of sought function *DIS(S)*.

Thus, the approach proposed here permits to transfer the whole difficulty of choosing specification of the forcing term in the physical space, $P_i(\mathbf{u}, \eta)$, to the choice of it in the spectral representation, $P_i(a_\mathbf{k})$.

### 2.4. General kind of the wave dissipation term in a spectral form

Now, return to initial system of equations (4)-(7) and rewrite it in the linear and potential approximations without any external force, excluding turbulent forcing $P_i(\mathbf{u}, \eta)$ introduced in the previous subsection. Accepting the following definitions

$$\mathbf{u}_w(\mathbf{x}, z, t) = \vec{\nabla}_3 \varphi(\mathbf{x}, z, t) \qquad , \qquad (25)$$

$$\Phi(\mathbf{x}, t) \equiv \varphi(\mathbf{x}, t)\big|_{z=\eta(\mathbf{x})} \qquad , \qquad (26)$$

one finds that two unknown functions, surface elevation field $\eta(\mathbf{x}, t)$ and velocity potential at the surface $\Phi(\mathbf{x}, t)$, are described by the following equations

$$\frac{\partial \Phi}{\partial t} + g\eta = -\hat{P}(\eta, \Phi), \qquad (27)$$

$$\frac{\partial \eta}{\partial t} = \frac{\partial \Phi}{\partial z}, \qquad (28)$$

$$\Delta \varphi = 0 \qquad \text{and} \qquad \frac{\partial \varphi}{\partial z}\big|_{z=-\infty} = 0. \qquad (29)$$

The first two equations are written at surface $\eta(\mathbf{x}, t)$, whilst two equations (29) done in a whole water layer. The last term in the r. h. s. of (27) means the result of transition to the potential representation for the turbulence forcing, i.e. $\hat{P}(\eta, \Phi) = (\vec{\nabla}_3)^{-1}[\mathbf{P}(\eta, \mathbf{u})]$. For more certainty, note that the first two equations, (27) and (28), are the main ones, as far as they determine the unknown fields, $\eta(\mathbf{x},t)$ and $\Phi(\mathbf{x},t)$. Equations (29) are auxiliary, they are used for determination of the vertical structure for potential $\varphi(\mathbf{x},z,t)$, only.

To make a transition into spectral representation, we introduce, as usually, the following Fourier-decompositions

$$\eta(\mathbf{x},t) = const \cdot \int\limits_{\mathbf{k}} \exp[i(\mathbf{k}\mathbf{x})]\eta_{\mathbf{k}}(t)d\mathbf{k} \ , \tag{30a}$$

$$\varphi(\mathbf{x},z,t) = const \cdot \int\limits_{\mathbf{k}} \exp[i(\mathbf{k}\mathbf{x})]f(z)\varphi_{\mathbf{k}}(t)d\mathbf{k} \ . \tag{30b}$$

Hereafter, the wave vector, $\mathbf{k}$, as well as the spatial vector, $\mathbf{x}$, has the horizontal components, only, and $f(z)$ is the so called vertical structure function, determined from Eqs. (29) solved analytically. In our case, $f(z)$=exp(-$kz$), what, by the way, dose not play any principal role later (see details, for example, in Hasselmann 1974, Polnikov 2007).

After substitution of representations (30) into system of Eqs. (27)-(29), equations (29) give the solution for the potential structure function mentioned above, and the other two equations get the kind

$$\dot{\Phi}_{\mathbf{k}} + g\eta_{\mathbf{k}} = -\Pi(\mathbf{k},\eta_{\mathbf{k}},\Phi_{\mathbf{k}}) \quad , \tag{31}$$

$$\dot{\eta}_{\mathbf{k}} = k\Phi_{\mathbf{k}} \quad . \tag{32}$$

Here, the point above wave variables means the partial derivative in time, and $\Pi(\mathbf{k},\eta_{\mathbf{k}},\Phi_{\mathbf{k}}) \equiv F^{-1}[\hat{P}(\eta,\Phi)]$ is the new denotation of forcing function where operator $F^{-1}$ means the inverse Fourier-transition applied to forcing function $\hat{P}(\eta,\Phi)$ (see technical details in the last two references).

System (31)-(32) is easily reduced to one equation having a sense of the well known equation for a harmonic oscillator with a forcing:

$$\ddot{\eta}_{\mathbf{k}} + gk\eta_{\mathbf{k}} = -k\Pi(\mathbf{k},\eta_{\mathbf{k}},\dot{\eta}_{\mathbf{k}}) \quad . \tag{33}$$

Solution of (33), written in the kind of evolution equation for a wave spectrum, can be carried out with the technique used in (Hasselmann 1974).

Following to this technique, we introduce the generalized spectral variables

$$a_\mathbf{k}^s = 0.5(\eta_\mathbf{k} + s\frac{i}{\sigma(k)}\dot{\eta}_\mathbf{k}), \quad \text{(where } s = \pm \text{ and } \sigma(k) = (gk)^{1/2}) , \quad (34)$$

and rewrite Eq. (33) in the kind

$$\dot{a}_\mathbf{k}^s + is\sigma(k)a_\mathbf{k}^s = -is\sigma(k)\Pi(\mathbf{k},\eta_\mathbf{k},\dot{\eta}_\mathbf{k})/2g . \quad (35)$$

Now, we accept the definition of the wave spectrum, used in (Hasselmann 1974)

$$2 << a_\mathbf{k}^s a_\mathbf{k}^{s'} >> = S(\mathbf{k})\delta(s+s') , \quad (36)$$

where the doubled brackets $<<.>>$ mean an averaging over the statistical ensemble for wind waves. To finish the evolution equation derivation, one needs to do the following steps:

(1) to multiply Eq. (35) by the complex conjugated component, $a_\mathbf{k}^{-s}$;

(2) to sum the newly obtained equation with the original one, (35);

(3) to make an ensemble averaging the resulting summarized equation.

Finally, one gets the most general evolution equation for a wave spectrum of the kind

$$\dot{S}(\mathbf{k},t) = \frac{2\sigma_k}{g}\text{Im} << \Pi(\mathbf{k},\eta_\mathbf{k},\dot{\eta}_\mathbf{k})a_\mathbf{k}^- >> \equiv -DIS(S) . \quad (37)$$

General kind of the sought dissipation term, $DIS(S)$, can be found after specification of the forcing function $\Pi(\mathbf{k},\eta_\mathbf{k},\dot{\eta}_\mathbf{k})$. To get the posed aim, it is sufficient to use the closure formula given by (24).

Due to qualitative feature of closure (24), there is no need to reproduce all mathematical procedures explicitly. To get the sought result: derivation of the dissipation term as a function of wave spectrum, $DIS(S)$, it is important only to take into account the following theoretical considerations:

(a) Structure of generalized variables (34) includes a sum of Fourier-components for elevation variable $\eta_\mathbf{k}$ and for velocity potential one $\dot{\eta}_\mathbf{k} \propto \Phi_\mathbf{k}$;

(b) Initial representation of forcing term (24) includes analogous sums for derivatives what means that the forcing term can be expressed via the generalized variables in the form

$$\Pi(\mathbf{k}, \eta_\mathbf{k}, \dot{\eta}_\mathbf{k}) = function(a_\mathbf{k}^s, a_\mathbf{k}^{-s}); \tag{38}$$

(c) Due to averaging over turbulent scales, the exponential phase factors in the Fourier-representation for $\Pi(\mathbf{k}, \eta_\mathbf{k}, \dot{\eta}_\mathbf{k})$ alike the form

$$\int\limits_{\mathbf{k}_3} d\mathbf{x} \left[ \exp(-i\mathbf{k}_3\mathbf{x}) < \lambda_i \lambda_j \int\limits_{\mathbf{k}_1} \int\limits_{\mathbf{k}_2} \phi(\mathbf{k}_1, \mathbf{k}_2) \exp(i\mathbf{k}_1\mathbf{x}) \exp(i\mathbf{k}_2\mathbf{x}) \eta_{\mathbf{k}_1} \eta_{\mathbf{k}_2} d\mathbf{k}_1 d\mathbf{k}_2 > \right] \tag{39}$$

may be simply omitted. It means that form (39) gets the kind

$$L_{ij}^2 \int\limits_{\mathbf{k}_1} \int\limits_{\mathbf{k}_2} \phi(\mathbf{k}_1, \mathbf{k}_2) \eta_{\mathbf{k}_1} \eta_{\mathbf{k}_2} d\mathbf{k}_1 d\mathbf{k}_2 \tag{40}$$

with statistical coefficient $L_{ij}$ as a constant of the theory. This assumption means that the averaging procedure over the turbulent scales suppresses the wave-like phase structure in the turbulent forcing term.

It needs to mention especially that newly introduced theoretical assumption (c) allows executing the inverse Fourier-transitions in the nonlinear summands of forcing term $P(\eta, \Phi)$ without appearance of residual integral-like convolutions containing the resonance-like factors for a set of three wave vectors, which are typical in the conservative nonlinear theory (see technical details, for example, in Krasitskii 1994, Polnikov 2007, Zakharov 1974). Thus, on basis of the considerations mentioned, it is quite reasonable (and sufficient for getting the aim posed) to represent the final expression for $\Pi(\mathbf{k}, \eta_\mathbf{k}, \dot{\eta}_\mathbf{k})$ in the most simple kind

$$\Pi(\mathbf{k},\eta_{\mathbf{k}},\dot{\eta}_{\mathbf{k}}) = \sum_{s_i,s_j} T_{ij}(\mathbf{k})a_{\mathbf{k}}^{s_i}a_{\mathbf{k}}^{s_j} \quad . \tag{41}$$

This form of function $\Pi(\mathbf{k},\eta_{\mathbf{k}},\dot{\eta}_{\mathbf{k}})$ retains the main feature of the forcing: nonlinearity in wave amplitudes $a_{\mathbf{k}}^{s}$. Herewith, both the explicit kind of multipliers $T_{ij}(\mathbf{k})$ and the certain representation of the quadratic form in the r. h. s. of (41) is not principle, as far as the main physical feature is retained.

Now, one can get the general kind of the r. h. s. in evolution equation (37), by using the procedure of multiplication, averaging, and so on, applied to Eq. (35) and described above in items (1)-(3). The first result of this procedure can be found by the following way.

Substitution of (41) into (37) results in a sum of the third statistical moments of the kind $L_{k,k,k}^{s1,s2,s3} \equiv <<a_{\mathbf{k}}^{s1}a_{\mathbf{k}}^{s2}a_{\mathbf{k}}^{s3}>>$ in the r. h. s. of (37):

$$\dot{S}(\mathbf{k}) = -i \sum_{s1,s2,s3} f_1(\mathbf{k},s1,s2,s3) T_{k,k,k}^{s1,s2,s3} L_{k,k,k}^{s1,s2,s3} \tag{42}$$

where $f_1(\ldots)$ is the intermediate function of the theory. Due to an even power in wave amplitudes for the wave spectrum (by definition 36), any third moment $L_{k,k,k}^{s1,s2,s3}$ can not be directly expressed via spectrum function $S(\mathbf{k})$. In such a case, according to a common technique of the nonlinear wave theory, one should use main equation (35) to write and solve auxiliary equations for each kind of the third moments, $<<a_{\mathbf{k}}^{s1}a_{\mathbf{k}}^{s2}a_{\mathbf{k}}^{s3}>>$, and to put these solutions into spectrum evolution equation (42). This procedure resembles a derivation of the three-wave kinetic evolution equation (see, for example, Zakharov 1974).

From the kind of the r. h. s. of Eq. (35), it is clear that any third moment could be expressed via a set of the fourth moments of the kind $<<a_{\mathbf{k}}^{s1}a_{\mathbf{k}}^{s2}a_{\mathbf{k}}^{s3}a_{\mathbf{k}}^{s4}>>$, having a lot of combinations for superscripts $s_i$. In cases then the condition $s1+s2+s3+s4 \neq 0$ is met, a part of fourth moments must be put a zero, according to definition (36). Residual fourth moments can be split into a sum of products of the second moments, $<< a_{\mathbf{k}}^{s1} a_{\mathbf{k}}^{-s1} >>$, each of which is expressed via spectrum $S(\mathbf{k})$ by definition (36). By this way the first nonvanishing spectral summand appears in the r.h.s. of evolution equation (37), and this summand is proportional to the second power in spectrum $S(\mathbf{k})$:

$$\dot{S}(\mathbf{k})4 = \mathrm{Re} \left\{ \sum_{s1,s2,s3} f_2(\mathbf{k}, s1, s2, s3) \left| T_{k,k,k}^{s1,s2,s3} \right|^2 FN_2[S(\mathbf{k})] \right\} + Res \qquad . \qquad (43)$$

Here, $f_2(\ldots)$ is the proper theoretical function, $FN_2[S(\mathbf{k})]$ is the functional of the second power in spectrum $S(\mathbf{k})$, and $Res$ are some residual moments of the fourth power in amplitudes $a_{\mathbf{k}}^{s}$.

The procedure described can be continued for a residual part of the fourth moments through the chain of actions described above what results in a sum of terms of the third power in spectrum in the r.h.s. of evolution equation (37). Eventually, the procedure mentioned provides for the power series in spectrum $S(\mathbf{k})$ in the r.h.s. of (37), starting from the quadratic term. As far as the whole r.h.s. of Eq. (37) has, by origin, a meaning of the dissipative evolution mechanism for a wave spectrum, the proposed theory results in function $DIS(S, \mathbf{k}, \mathbf{W})$ of the following general kind:

$$DIS(S, \mathbf{k}, \mathbf{W}) = \sum_{n=2}^{N} c_n(\mathbf{k}, \mathbf{W}) FN_n[S(\mathbf{k})]. \qquad (44)$$

Specification of coefficients $c_n(\ldots)$, including their dependence on the wave-origin factors, and determination of the final value of $N$ in series (44), is based on principles which are not related to hydrodynamic equations. These points will be specified below by a separate way.

As a conclusion of this section, it is worth while to emphasize that the main fundamental of the theory, providing for result (44), is nothing else as the nonlinear feature of the Reynolds stress closure justified physically in subsection *2.3*. Consequently, the nonlinear feature of result (44) is justified as well.

## 3. Parameterization of the dissipation term and its properties

In this section, using an ideology of earlier papers by the author (Polnikov 1995, 2005), we consider the following points:

(a) Certain parameterization of dissipation term *DIS(S ,*$\mathbf{k}$*,*$\mathbf{W}$*)* given in the kind (44);

(b) Physical meaning of the certain parameters of *DIS(S ,*$\mathbf{k}$*,*$\mathbf{W}$*)*, finally introduced;

(c) Correspondence of the parameterization for *DIS(S ,*$\mathbf{k}$*,*$\mathbf{W}$*)* to experimental effects E1-E4 mentioned in Sect.1;

(d) Evidence of effectiveness of the proposed version for *DIS(S ,*$\mathbf{k}$*,*$\mathbf{W}$*)*.

### 3.1. *Specification of function DIS(S ,*$\mathbf{k}$*,*$\mathbf{W}$*)*

First of all, here we accept the simplest representation of (44), supposing that each functional $FN_n[S(\mathbf{k})]$ is a local function of $S(\mathbf{k})$, i.e. $FN_n[S(\mathbf{k})] \propto S^n(\mathbf{k})$ [1]. In such a case, it is very easy to estimate the value of power $N$, which can limit the general representation of *DIS(S,*$\mathbf{k}$*,*$\mathbf{W}$*)* in the kind of series (44). To do it, let us use the following well known fact of existence of a stable and equilibrium spectral shape, $S_{eq}(\sigma)$, usually attributed to a fully developed sea (Komen et al. 1994). Not addressing to discussion about a falling law for the tail part of the equilibrium wave spectrum, accept here that in the frequency domain given by the following condition

$$\sigma > 2.5\sigma_p \qquad , \qquad (45)$$

---

[1] This representation is based on the assumption of suppression the wave-like phase structure in the turbulent forcing terms. In a general case, functional $FN_n[S(\mathbf{k})]$ should include a series of integral-like convolutions of the $n$-th power in $S(\mathbf{k})$. Consideration of this point is out of the present paper, it is rather related to a future elaboration of the theory (see Sec. 4).

where $\sigma_p$ is the peak frequency of the spectrum $S(\sigma,\theta)$, the equilibrium spectrum has the shape

$$S_{eq}(\sigma) = \alpha_p \, g^2 \sigma^{-5} \tag{46}$$

corresponding to the standard Phillips' spectrum (Phillips, 1958). This allows us to introduce the small parameter, $\alpha$, defined in the whole frequency band by the form[2]

$$\alpha = \max[S(\sigma,\theta)\sigma^5 / g^2] << 1 \qquad (0 < \sigma < \infty). \tag{47}$$

It is worth while to mention that parameter $\alpha$ defined by (47) has the order of the second power of mean wave slope, i.e. it is really quite small ($\alpha \approx \alpha_p \approx 0.01$).

Existence of small parameter in a spectral representation for any dynamical characteristic of wind-wave field means that its spectral series is a series in a small parameter. Hereof, with no lose of theoretical accuracy, one should immediately conclude that series (44) can be restricted by the first term, i.e $N = 2$. Hence, after some algebra in terms of the similarity (scaling) approach, the dissipation function can be written in the form

$$DIS(...) \cong c_2(...)S^2(\mathbf{k}) = \gamma(\sigma,\theta,\mathbf{W})\frac{\sigma^6}{g^2}S^2(\sigma,\theta) \tag{48}$$

where the unknown dimensional factor, $c_2$, is changed by the unknown dimensionless function, $\gamma(\sigma,\theta,\mathbf{W})$, all arguments of which a written explicitly. Besides, in the r.h.s. of (48), all powers of frequency $\sigma$ and spectrum $S$ are brought together. It is left to define an explicit expression for dimensionless function $\gamma(\sigma,\theta,\mathbf{W})$ by means of similarity approach.

General kind of $\gamma(\sigma,\theta,\mathbf{W})$ can be easily defined on the basis of assumption for existence of equilibrium spectrum in a frequency domain (45), accepted before. By definition, at the tail of wave spectrum, a spectral equilibrium means that the balance of terms in source function $F$ is close to zero for a fixed spectrum shape $S_{eq}(\sigma,\theta)$, i.e.

---

[2] To spread an existence of small parameter to representation of theoretical spectrum $S(\mathbf{k})$, one should taken into consideration the chain of dimensional ratios: $S(\sigma) \propto S(\sigma,\theta) \propto S(\mathbf{k})k(\partial \sigma / \partial k)^{-1}$.

$$F\big|_{S=S_{eq}} = \big[NL + IN - DIS\big]\big|_{S=S_{eq}} \approx 0 \,. \tag{49}$$

Now, take into account that in high frequency domain (45), a relative contribution of the nonlinear term *NL* to the source function is less than 10%. Then, ratio (49) gets the form

$$\big[IN - DIS\big]\big|_{S=S_{eq}} \approx 0 \,. \tag{50}$$

Accepting ratio (9) as the basic parameterization for input term *IN*, by using ratios (48) and (50) one can easily find a formal expression for function $\gamma(\sigma, \theta, \mathbf{W})$, eventually resulting in the following expression for the dissipation term

$$DIS(\sigma, \theta, S, \mathbf{W})\big|_{S=S_{eq}} \approx \beta(\sigma, \theta, \mathbf{W}) \frac{\sigma^6}{g^2} S^2(\sigma, \theta) \qquad , \tag{51}$$

which is valid in the spectrum tail domain corresponding to ineqality (45).

To get a final specification of $DIS(\sigma, \theta, S, \mathbf{W})$, valid in the whole frequency domain, one should take into account the following empirical features:

(a) Specific character of dissipation processes in the energy containing domain, i.e. in the vicinity of the spectral peak where $\sigma \approx \sigma_p$ (empirical effects E2, E3);

(b) Inevitable "background" dissipation existing when $\beta(\sigma, \theta, \mathbf{W}) \le 0$;

(c) Two-lobe feature of the angular spreading function, $T(\sigma, \theta, \theta_w)$ (empirical effect E4).

With account of the said above, finally we have the following specification

$$DIS(\sigma, \theta, S, \mathbf{W}) = c(\sigma, \theta, \sigma_p) \max\big[\beta_L, \beta(\sigma, \theta, \mathbf{W})\big] \frac{\sigma^6}{g^2} S^2(\sigma, \theta) \qquad . \tag{52a}$$

Here, the known dimentionless function $\beta(\sigma, \theta, \mathbf{W})$ is given by a certain empirical formula, the kind of which is not principal at the moment (for more details, see Polnikov 2005); $\beta_L$ is the "background" dissipation parameter, the default value of which is $\beta_L = 0.00005$; and $c(\sigma, \theta, \sigma_p)$ is the dimensionless fitting function describing peculiarities of the dissipation rate in the vicinity of spectrum peak frequency $\sigma_p$. According to Polnikov (2005), $c(\sigma, \theta, \sigma_p)$ could be given by the following phenomenological formula

$$c(\sigma, \theta, \sigma_p) = C_{dis} \max\left[ 0, \ (1 - c_\sigma(\sigma_p / \sigma)) \right] T(\sigma, \theta, \sigma_p) \qquad (52b)$$

where the angular spreading factor is accepted in the form

$$T(\sigma, \theta, \theta_w, \sigma_p) = \left\{ 1 + 4\frac{\sigma}{\sigma_p}\sin^2(\frac{\theta - \theta_w}{2}) \right\} \max\left[ 1, \ 1 - \cos(\theta - \theta_w) \right], \qquad (52c)$$

$C_{dis}$ and $c_\sigma$ are the fitting parameters, $\theta_w$ is the mean wind direction, and the standard designation, max[$a$, $b$], means a choice of maximum among two values under the brackets.

Hereby, the sought parameterization of $DIS(\sigma, \theta, S, \mathbf{W})$ is totally defined, and a general theoretical justification for the dissipation term is finished. It is left to add that in the course of specification of function $\gamma(\sigma, \theta, \mathbf{W})$, one has a certain arbitrariness related to a choice of the equilibrium spectrum shape, $S_{eq}(\sigma, \theta, \mathbf{W})$, and the kind for the angular form, $T(\sigma, \theta, \theta_w)$, as functions of their arguments. This arbitrariness is justified to some extent by uncertainty of the proper functions obtained experimentally (Rodrigues and Soares 1999, Young and Babanin 2006). Nevertheless, the general kind of parameterization (48) is robust to the uncertainties of such a kind. Just this circumstance allows hoping on universality of its application in different numerical models for wind waves.

*3.2. Physical meaning of the dissipation term parameters and correspondence to empirics*

For completeness of the theory, it is important to reveal physical meaning of all parameters which are used in the proposed version of function $DIS(\sigma, \theta, S, \mathbf{W})$, given by formulas (52). To say nothing about a fully phenomenological angular-spreading function $T(\sigma, \theta, \theta_w)$, note that the theory has only three fitting parameters: $C_{dis}$, $c_\sigma$ and $\beta_L$.

Meaning of coefficient $C_{dis}$ is evident and simple. It regulates the dissipation intensity. This parameter is inevitable in any approach to the problem of source function construction. Moreover, just $C_{dis}$ is strongly varied while fitting any numerical model of the kind (1), having representation of the source function as a sum of several separate evolution mechanisms.

Meaning of parameter $c_\sigma$ consists, mainly, in separation of dissipation features in too frequency domains: the vicinity of spectral peak and the spectrum tail domain. In the latter domain, the role of $c_\sigma$ becomes negligible. But in the vicinity of spectral peak, $c_\sigma$ regulates an extent of suppression of the "pure turbulent" dissipation intensity described by ratio (51). Here one can see manifestation of empirical effects E2 and E3 mentioned in Sec.1.

Really, variation of parameter $c_\sigma$ does change radically relative intensity of the dissipation rate in both frequency domains mentioned above (an analogue of effects E2, E3), what result in different variations for different mean wave characteristics (the significant wave height, $H_S$, the dominant period, $T_p \propto \sigma_p^{-1}$, and the mean wave period, $T_m$). For example, decreasing $c_\sigma$ results in a lowering the rate of the spectrum peak growth in course of wave evolution and, consequently, increasing a relative contribution of the spectrum tail part to the value of mean period $T_m$ estimated by the formula

$$T_m = \frac{2\pi \int \sigma^{-1} S(\sigma) d\sigma}{\int S(\sigma) d\sigma} \qquad . \qquad (53)$$

With a glance that a value of dominant frequency $\sigma_p$ is mainly defined by the nonlinear mechanism of evolution (Komen et al. 1994, Polnikov 2005, 2009) and weakly depend on a value of $c_\sigma$, decrease of $c_\sigma$ provides for reducing mean wave period $T_m$ and significant wave height $H_S$, without remarkable change of dominant period $T_p \propto \sigma_p^{-1}$. Naturally, an increase of $c_\sigma$ results in the inverse effect. This feature of parameter $c_\sigma$ was effectively used in executing tasks of fitting and verification of new source function (Polnikov at al. 2008, Polnikov and Innocentini 2008).

Finally, we should say some words about a meaning of parameter $\beta_L$. Its main role is to regulate a wave dissipation rate at time moments of sharp changing the local wind (falling or turning). It is clear that at these time moments, a value of increment $\beta(\sigma, \theta, \mathbf{W})$, corresponding to the former wind direction, is radically reduced, resulting in reducing a rate of breaking for the wave components running along the former wind direction. But in reality, certain, background turbulence retains, as far as it "lives" in the water upper layer for a long time. Thus, the background turbulence should provide remarkable attenuation of the wave components running in the former wind direction, which became now a swell. The said explains both a meaning of parameter $\beta_L$ and the role of background dissipation in general. An initial choice of value for $\beta_L$ is based on some empirical and theoretical estimations (see references in Polnikov 2005). But the final value should be found during the fitting a whole numerical model, i.e. in a concert with a choice of all others parameters used in a source function.

The joint dynamics of the input term and the dissipation term including a background constituent gives raise a quicker wind sea accommodation to a new wind direction. Just this effect is less described by numerical models with traditional dissipation term without background constituent (WAM or WW), what was explicitly shown in proper papers (Polnikov at al. 2008, Polnikov and Innocentini 2008) (see Fig. 2 below).

Thus, the said above allows stating that all fitting parameters introduced into the proposed version of dissipation term function $DIS(\sigma, \theta, S, \mathbf{W})$ has both the special-purpose and the physical meaning features.

Additionally, it is worth while to emphasize the important role of quadratic power of function *DIS (S)* in spectrum *S*, allowing easy regulating the modeled dependence of the equilibrium spectrum, $S_{eq}(\sigma,\,\theta)$, on frequency $\sigma$ by means of varying a frequency power in ratio (51). It is caused by the fact that an expression for an equilibrium spectrum shape follows directly from the commonly accepted balance condition (50). Really, substituting the linear in spectrum input term *IN(S)* and quadratic in spectrum dissipation term *DIS(S)* into Eq. (51) gives simply an equation for a shape of spectrum $S_{eq}(\sigma,\,\theta)$. Thus, the theory has no restrictions for the shape of equilibrium spectrum.

Particularly, if one wants to postulate the equilibrium spectrum of Toba's shape in the kind (Komen et al 1994)

$$S_{eq,T}(\sigma) = \alpha_T u_* \, g\sigma^{-4}, \tag{54}$$

he should separately extract dimensionless multiplier $g/u_*\sigma$ from function $\gamma(\sigma,\theta,\mathbf{W})$ in (48), which should be saved in the r.h.s. of ratios (51) and (52a). In such a case, the balance condition (48) results in the sought equilibrium spectrum of the kind (54).[3]

The shown encouraging comparison of present theoretical version for dissipation term to the well established empirical effects mentioned in Sec.1 could be finished by the evident correspondence of the accepted angular spreading function of the kind (52c) to the recently revealed empirical fact: effect E4 found in Young and Babanin 2006.

Herewith, regarding to the threshold feature of breaking phenomenon (effect E1), it is easily understood that, in a spectral representation for the dissipation term, this effect is smoothed due to statistical distribution of breaking events in a stochastic wind sea field. Thus, effect E1 is hardly appropriate to a real feature of the turbulence dissipation approach.

---

[3] Note that by this way one changes the dependence of *DIS* on wind W. Thus, the way shown here could be used to get the best balance between input and dissipation term as functions of the wind (if anybody knows this dependence for equilibrium spectrum).

### 3.3.  Proofs of effectiveness of new dissipation function

The only convincing way to prove a superiority of new dissipation function, in the aspect of solving numerical simulation tasks for wind waves, is to carry out the so called procedure of "comparative verification". According to papers (Polnikov at al. 2008, Polnikov & Innocentini 2008), regulations of comparative verification procedure demand fulfillment of the following series of conditions:

(a)  Reasonable data base of reliable observations for wind waves;

(b)  Reliable wind field given on a rather thick space-time grid for the whole period of observations;

(c)  Well designed mathematical part for numerical  model of the kind (1);

(d)  Choice of a widely recognized wind wave model as a reference (etalon) for comparison.

Just these conditions were realized in the papers mentioned above where the models WAM and WW were used as etalons. Herewith, due to specificity of the wave data available, the comparative verification was done by means of comparing the main integral wave characteristics: significant wave height, $H_s$, and mean wave period, $T_m$, instead of two-dimensional wave spectra. The comparison was carried out between numerical results obtained with the original models, WAM and WW, and analogous results done with modifications of the both models, realized by means of replacing the source functions, only. The role of new source function was attributed to the function proposed in (Polnikov 2005) with the *DIS(S)* described above in subsection *3.1*.

In both cases, the modification of source functions consists in replacing the original terms *NL* and *IN*, used in WAM and WW, by new terms represented in the forms maintaining the physics enclosed in them. But the modification of original terms *DIS* was done by *DIS(S)* given in accordance to the version described by formulas (52), changing the physics involved radically. Therefore, the whole deference in accuracy of these calculations was ascribed to changing the term *DIS* in the modified source function. So, in the case of accuracy enhancement during comparative verification, the said permits to say about effectiveness of new version for *DIS*, and vice-verse.

In both papers referred above, it was convincingly shown that replacing the term *DIS* results in reducing the root-mean-square error for significant wave height $H_s$ on 15-20%, and it does more then on 20-30% for mean wave period $T_m$. For more completeness of cogency, the time history for significant wave height $H_s(t)$ is shown in Fig. 2. It permits to check visually that replacing the dissipation term gives a better description just for extreme waves and for wave decay time moments (ascribed to the moments of local wind change at the ocean point under consideration , i.e. the buoy location point).

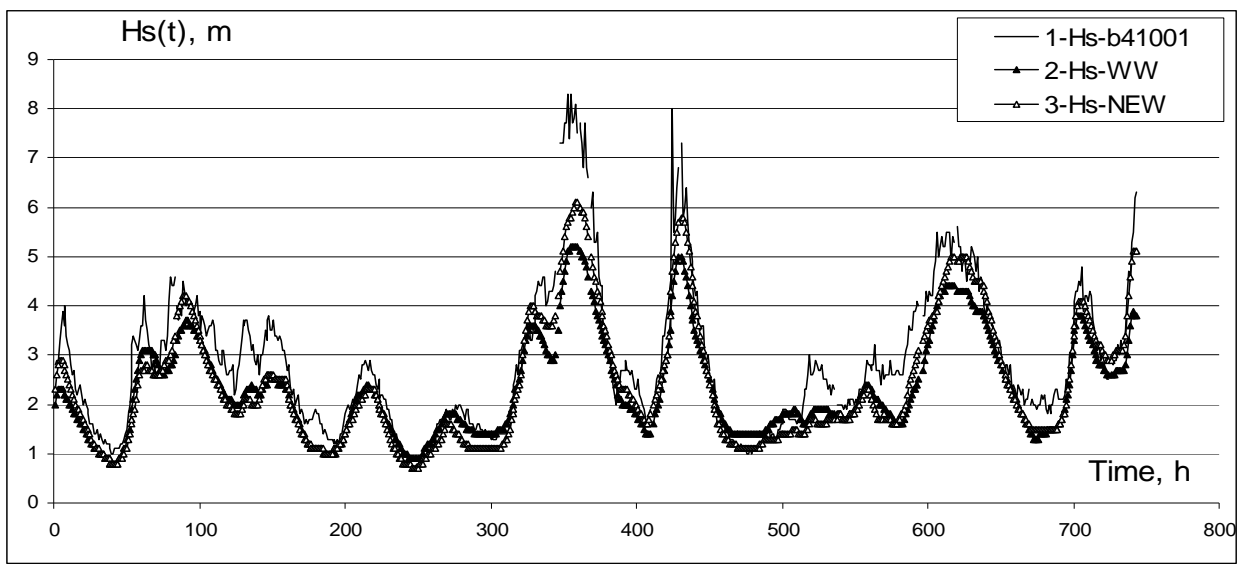

Fig. 2. The time history of significant wave height $H_s(t)$ at the location point of buoy 41001 in the North Atlantic for the period January 2006: 1- buoy observations, 2 – results of modelling with the model WW, 3 – results of modelling with the modified model WW (following to Polnikov and Innocentini 2008).

Here we have no place for full analysis of the comparison methodic and for demonstrating more details, which can be found in the papers referred. At the moment it is important to emphasize the fact of remarkable increasing accuracy of simulations, achieved with the well known models, WAM and WW, due to changing their physical content (i.e. source function). Herewith, the key item of this changing is the dissipation term, detailed physical justification of which is the main aim of this work.

### 4. Conclusion

In conclusion of the paper, we dwell shortly on the fundamentals of the theory proposed, domain of the theory applicability, and prospects of its further elaboration.

Firstly, about fundamentals. The most general among them is the statement that on the scales of the wind wave spectrum evolution, corresponding to equation (1) (i.e. hundreds of dominant wave periods), the general physical reason for wave dissipation is the turbulence in a water upper layer, induced by the whole aggregate of dissipative processes at the air-sea interface. Naturally, the laws of such a turbulence formation are hardly known for us. Therefore, the following assumptions are attracted to constructing the theory:

(1)   Nonlinear feature of Reynolds stresses resulting in the wave dissipation;

(2)   Destruction of the wave-like phase structure for the wave variables which are involved to a closure of Reynolds stresses.

Reasonableness of the first assumption is rather evident (see subsection *2.2* above), meanwhile the second one needs some additional justification.

The most essential aspect of this justification consists in a physically clear idea that Reynolds stress is a statistical characteristic, for which the wave-like features of motion are not appropriate due to averaging over turbulent scales. Therefore, while making closure of this characteristic via the large scale motions (in our case – wave motions), one have some grounds to introduce the assumption under discussion. The main profit of this assumption is a simplification of further algebra manipulations. Eventually, just assumption (2) results in the local structure of the final dissipation term in the wave vectors space, i.e. the structure without any integral-like convolution summands.

Acceptability and validity of the assumption, provided by the need to get final formulas in a simple and practically applicable form, can be checked a posteriori, i.e. by means of verification procedure described in subsection *3.3* above.

Secondly, about domain of applicability. Due to quadratic power in spectrum for the dissipation term, the question does raise dealing with the time scales of wave dissipation. It is clear that each power in spectrum for any term in the source function enlarges the time scale of its action. Therefore, in our version of source function, the input mechanism is the quickest one. It is balanced by the more slow dissipation term which, in the high frequencies domain (45), plays the role of a spectrum tail smoother. (Here we pay attention that each individual breaking event does not correspond to this time scale). Farther, on a more large time scale, the most slow nonlinear evolution mechanism starts to play a remarkable role, resulting in the shift of spectrum as a whole to the lower frequency domain (Cavaleri et al. 2007, Komen et al. 1994, Polnikov 2009).

In such a case, the question of joint treatment of very quick breaking processes and slow wave dissipation process does arise. The only logical reconciliation of this "apparent contradiction" is our statement that the breaking event, itself, is not the fact of wave energy dissipation. Moreover, the well known Hasselmann's hypotheses about "weak in mean", accepted in this theory as well, means, in our mind, that any breaking event is a certain, strongly non-conservative and nonlinear disturbance redistributing wave energy through the whole wave spectrum. Herewith, one may even expect a short-time increase of intensity for some sideband wave components during the local action of breaking event, as it is seen in Fig. 1.

Being a quick process, breaking is out of spectral consideration when one says about a loss of wave energy on the scales of hundreds dominant periods. From this point of view, it becomes evident that the effect of threshold-feature for breaking events, observed experimentally (Banner and Tian 1998, Young and Babanin 2006) and numerically (Zakharov at al. 2007) is smoothed in a spectral representation for dissipation rate of wave spectrum due to statistics of breaking on the scales taken into the consideration. This treatment reconciles the quick and strong nonlinear breaking events with the slow action of dissipation term in a spectral representation.

Thus, the said above defines a certain domain for applicability of the proposed theory: it does not describe individual quick and random dissipation events, being valid for description of wave spectra evolution realized on the scales of variability for a whole statistical ensemble of a random wave field.

The last, about prospects of further elaboration of the theory. Further development of the theory, in our mind, lies in the direction of elaborating the assumption about destruction of the wave-like phase structure for variables involved in a closure of Reynolds stresses. It could be at some extent reasonable to suppose that, in reality, some parts of the phase factors are conserved in the Fourier-decompositions for the wave variables under the brackets of turbulent scales averaging (see Eq. 39). Then, in the course of derivation of the final expression for $DIS$, some convolution expressions, including integrals over the wave vector $\mathbf{k}$-space (or the frequency space), can appear in the r.h.s. of Eq. (37). The latter should result in a radical change of a functional representation for the dissipation term, right up to appearing cumulative summands in $DIS(S)$, empirically introduced in Young and Babanin (2006). In the case of such a representation for $DIS$ term, the problem will rise, dealing with acceptability the balance condition of the form (50). This point needs more detailed investigation in future.

It seems that after wide discussion and verification of the results presented here, basing on existent data and on especially gathered in future, development of new versions of *DIS(S)* would be the next step in a further elaboration of the proposed theory for dissipation mechanism of wind waves.

### Acknowledgements

The author is grateful to Babanin A.V. for numerous discussions of the point, stimulating this paper. I acknowledge an interest of academicians of RAS Golitzin G.S. and Dymnikov V.P. to the problem. The work was supported in part by the Russian Federal Program "World Ocean" (State contract #6) and by the Russian Fund for Basic Research, # 08-05-13524-ophi-c.